\documentclass[pra,aps,twocolumn,nopacs,superscriptaddress,nofootinbib]{revtex4}

\usepackage{graphicx}  % Include  files
\usepackage{dcolumn}  % align table col. on decimal point
\usepackage{bm}           % bold math
\usepackage{bbm}
\usepackage{amsmath}
\usepackage{amssymb}
\usepackage{amsfonts}
\usepackage{enumerate}
\usepackage{indentfirst}
\usepackage{psfrag}
\usepackage{float}
\usepackage{color}

\def\ii{{\rm i}}  \def\ee{{\rm e}}

\def\xx{\hat{\bf x}}    \def\zz{\hat{\bf z}}

  \def\kpar{k_\parallel}  
    
\def\Eb{{\bf E}}

\def\lamp{\lambda_{\rm p}}    \def\Lp{L_{\rm p}}    \def\wp{\omega_{\rm p}}
  
\begin{document}
\title{Plasmonics in Atomically-Thin Crystalline Silver Films}

\author{Zakaria~M.~Abd~El-Fattah}
\thanks{These two authors contributed equally to the work.}
\affiliation{ICFO-Institut de Ciencies Fotoniques, The Barcelona Institute of Science and Technology, 08860 Castelldefels (Barcelona), Spain}
\affiliation{Physics Department, Faculty of Science, Al-Azhar University, Nasr City, E-11884 Cairo, Egypt}
\author{Vahagn~Mkhitaryan}
\thanks{These two authors contributed equally to the work.}
\affiliation{ICFO-Institut de Ciencies Fotoniques, The Barcelona Institute of Science and Technology, 08860 Castelldefels (Barcelona), Spain}
\author{Jens~Brede}
\affiliation{Donostia International Physics Center, Paseo Manuel Lardiza􏰀bal 4, 20018 Donostia-San Sebasti\'a􏰀n, Spain}
\author{Laura~Fern\'andez}
\affiliation{Centro de F􏰀\'{\i}sica de Materiales CSIC-UPV/EHU and Materials Physics Center, 20018 San Sebasti\'a􏰀n, Spain}
\author{Cheng~Li}
\affiliation{Department of Electrical Engineering, Yale University, New Haven, Connecticut 06511, USA}
\author{Qiushi~Guo}
\affiliation{Department of Electrical Engineering, Yale University, New Haven, Connecticut 06511, USA}
\author{Arnab~Ghosh}
\affiliation{Faculty of Engineering, Bar Ilan University, Ramat Gan – 5290002, Israel}
\author{A.~Rodr\'{\i}guez~Echarri}
\affiliation{ICFO-Institut de Ciencies Fotoniques, The Barcelona Institute of Science and Technology, 08860 Castelldefels (Barcelona), Spain}
\author{Doron~Naveh}
\affiliation{Faculty of Engineering, Bar Ilan University, Ramat Gan – 5290002, Israel}
\author{Fengnian~Xia}
\affiliation{Department of Electrical Engineering, Yale University, New Haven, Connecticut 06511, USA}
\author{J.~Enrique~Ortega}
\email{enrique.ortega@ehu.es}
\affiliation{Donostia International Physics Center, Paseo Manuel Lardiza􏰀bal 4, 20018 Donostia-San Sebasti\'a􏰀n, Spain}
\affiliation{Centro de F􏰀\'{\i}sica de Materiales CSIC-UPV/EHU and Materials Physics Center, 20018 San Sebasti\'a􏰀n, Spain}
\affiliation{Departamento de F􏰀\'{\i}sica Aplicada I, Universidad del Pa􏰀\'{\i}s Vasco, 20018 San Sebasti\'a􏰀n, Spain}
\author{F.~Javier~Garc\'{\i}a~de~Abajo}
\email{javier.garciadeabajo@icfo.es}
\affiliation{ICFO-Institut de Ciencies Fotoniques, The Barcelona Institute of Science and Technology, 08860 Castelldefels (Barcelona), Spain}
\affiliation{ICREA-Instituci\'o Catalana de Recerca i Estudis Avan\c{c}ats, Passeig Llu\'{\i}s Companys 23, 08010 Barcelona, Spain}

%\date{\today}

\begin{abstract}
Light-matter interaction at the atomic scale rules fundamental phenomena such as photoemission and lasing, while enabling basic everyday technologies, including photovoltaics and optical communications. In this context, plasmons --the collective electron oscillations in conducting materials-- are important because they allow manipulating optical fields at the nanoscale. The advent of graphene and other two-dimensional crystals has pushed plasmons down to genuinely atomic dimensions, displaying appealing properties such as a large electrical tunability. However, plasmons in these materials are either too broad or lying at low frequencies, well below the technologically relevant near-infrared regime. Here we demonstrate sharp near-infrared plasmons in lithographically-patterned wafer-scale atomically-thin silver crystalline films. Our measured optical spectra reveal narrow plasmons (quality factor $\sim4$), further supported by a low sheet resistance comparable to bulk metal in few-atomic-layer silver films down to seven Ag(111) monolayers. Good crystal quality and plasmon narrowness are obtained despite the addition of a thin passivating dielectric, which renders our samples resilient to ambient conditions. The observation of spectrally sharp and strongly confined plasmons in atomically thin silver holds great potential for electro-optical modulation and optical sensing applications.
\end{abstract}

\maketitle

%------------------------------------------------------
\section{INTRODUCTION}

The control of light at the nanoscale is a research frontier with applications in areas as diverse as biosensing \cite{AHL08,paper256}, optoelectronics \cite{MS16}, nonlinear optics \cite{DN07,SK16}, quantum optics \cite{CSH06,FLK14}, and nanorobotics \cite{ZDL15}. Metallic nanostructures play a pivotal role in this context because they host collective electron oscillations, known as plasmons, which can interact strongly with light. This enables a large confinement of optical energy down to nanometer-sized regions, thereby enhancing the associated electromagnetic fields by several orders of magnitude relative to externally incident fields \cite{LSB03}. Such appealing properties and the pursue of the noted applications have fueled intense research work into plasmonics to better understand and control these collective electronic excitations and cover a broad spectral range from the ultraviolet to the terahertz regimes. Progress has mainly relied on advances in nanofabrication and colloid chemistry, which allow producing engineered metallic nanostructures with on-demand plasmonic response \cite{NLO09,FWB10}.

Plasmons in atomic-scale systems have emerged as a source of extraordinary properties resulting from the fact that they are sustained by a comparatively small number of charge carriers. Electron energy-loss spectroscopy has been instrumental in revealing plasmons in systems such as C$_{60}$ molecules \cite{KC92}, carbon and boron-nitride single-wall nanotubes \cite{STK02,ASK05}, atomic gold wires grown on vicinal silicon surfaces \cite{NYI06}, few-atomic-layer silver films \cite{MRH99}, monolayer DySi$_2$ \cite{RNP08}, ultrathin indium \cite{CKH10} and silicide \cite{RTP10} wires, and graphene \cite{ZLN12}. Additionally, ultrathin TiN films have been demonstrated for refractory plasmonics \cite{GBS14,SRK17}, which contribute to configure the emerging field of transdimentional photonics \cite{BS19}. Among these materials, high-quality graphene has been found to sustain low-energy plasmons when it is highly doped, exhibiting large electro-optical tunability \cite{FRA12,paper196}, long lifetime \cite{NMS18}, and strong confinement compared with conventional plasmonic metals \cite{paper283,AND18}. Topological insulators \cite{DOL13} and black phosphorous \cite{HMP17}  have also been shown to display two-dimensional (2D) plasmons. Unfortunately, unlike noble-metal structures, the plasmons reported in these systems are either rather broad or lying at mid-infrared or lower frequencies, far from the technologically appealing near-infrared (NIR) regime. As a potential solution to this problem, electrochemically tunable plasmons have been revealed through optical spectroscopy in small polycyclic aromatic hydrocarbons \cite{paper215,paper260}, although their integration in fast commutation devices remains a challenge.

Atomically thin noble-metal films appear as a viable solution to achieve large electro-optical tunability \cite{paper236,paper277} within the NIR spectral range. However, crystalline quality is required to lower optical losses to the promised level for these materials in the plasmonic spectral region. Indeed, the presence of multiple facets in few-nanometer nanoparticles \cite{KV95,SKD12} and sputtered films \cite{paper326} produce broad plasmons characterized by a quality factor ($Q=$\,ratio of peak frequency to spectral width) of the order of $\sim1$, which averts their use in cutting-edge plasmonic applications.

Here, we report on the fabrication and the excellent plasmonic and electrical properties of wafer-scale atomically-thin crystalline silver films composed of only a few atomic layers. We use advanced surface-science techniques to fabricate and characterize Ag(111) films consisting of 7-20 atomic monolayers (MLs) on a clean Si(111) substrate, which we then cover with $\sim1.5$\,nm of Si in order to passivate them from air. The high atomic quality of the samples, which we confirm through scanning tunneling microscopy (STM), angle-resolved photoelectron spectroscopy (ARPES), high-resolution transmission electron microscopy (HRTEM), and low-energy electron diffraction (LEED), allows us to resolve sharp electronic vertical quantum-well states (QWs) and measure very low sheet resistances for thin films down to 7\,ML Ag(111) ($1.65\,$nm thick, $\sim20\,\Omega/$sq, just a factor of 2 higher than the bulk estimate). We obtain spectral evidence of confined plasmons by using electron-beam (e-beam) nanolithography to pattern ribbons on the silver films, resulting in measured plasmons with quality factors nearing $Q\sim4$ for 10\,ML ($\approx2.4$\,nm) films. These results reveal the ability of laterally-patterned few-atomic-layer atomically-flat silver to confine plasmons with similar lifetimes as bulk silver, thus extending 2D plasmonics into the technologically appealing NIR regime.

\begin{figure*}
\begin{centering}
\includegraphics[width=1\textwidth]{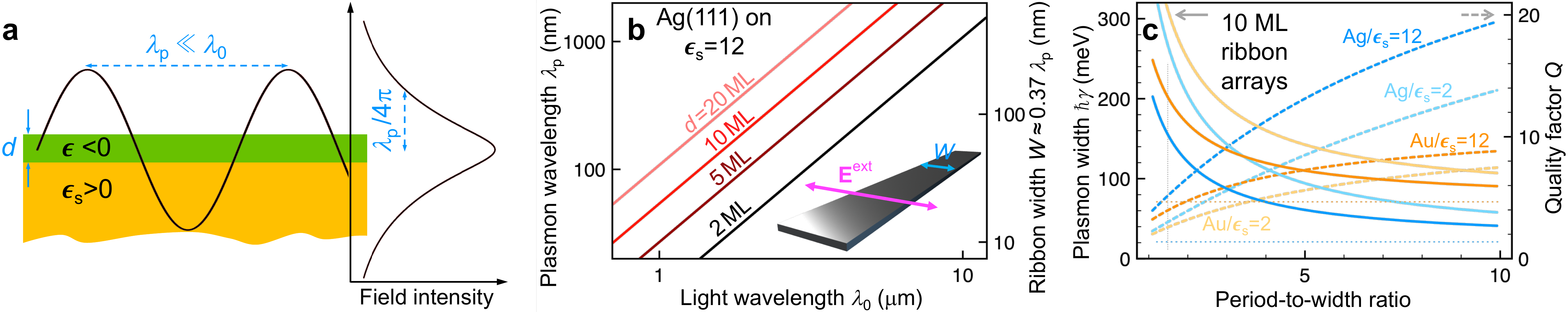}
\end{centering}
\caption{{\bf Properties of plasmons in atomically-thin metal films.} {\bf (a)} The plasmon wavelength $\lamp$ is small compared with the light wavelength $\lambda_0$, while the associated electric field extends a distance $\lamp/4\pi$ away from the film (for $1/e$ drop in field intensity), symmetrically on both sides of the interface (see Methods), regardless of dielectric environment and film composition. The sketch shows a cross section of an extended film (permittivity $\epsilon<0$) and substrate (permittivity $\epsilon_{\rm s}>0$) in a plane perpendicular to the surface, along with the in-plane harmonic oscillation of the plasmon field (sine profile) and exponential out-of-plane decay of its intensity (right plot). {\bf (b)} The plasmon wavelength scales linearly with metal thickness $d$ and quadratically with $\lambda_0$ as $\lamp=d\,(\lambda_0/L_1)^2$, where $L_1$ depends on the choice of materials and is rather large ($L_1\approx205\,$nm) for Ag on Si. A ribbon of width $W$ exhibits transverse dipolar resonances ({\it i.e.}, with in-plane polarization across the ribbon) determined by $W\approx0.37\,\lamp$. {\bf (c)} In ribbon arrays, the plasmon width has a radiative component that scales linearly with both the metal thickness and the inverse of the period-to-width ratio (see Methods), and depends on the choice of metal and substrate permittivity (see labels), therefore affecting the quality factor $Q$ as shown here for 10\,ML metal at $\lambda_0=1.55\,\mu$m wavelength. Dotted horizontal lines denote the long period limit for Ag and Au. The dotted vertical line shows the ratio used in this work.} \label{Fig1}
\end{figure*}

Like in graphene \cite{paper235}, metal films of small thickness $d$ in the few atomic-layer range allow us to dramatically reduce the in-plane surface-plasmon wavelength $\lamp$. In the Drude model (see Methods), we find $\lamp$ to scale linearly with $d$ and quadratically with the light wavelength $\lambda_0$ as
\begin{align}
\lamp=d\,\frac{\lambda_0^2}{L_1^2}
\label{lpl0}
\end{align}
(see Fig.\ \ref{Fig1}b), where $L_1$ is a characteristic length that depends on the combination of metal and substrate materials ({\it e.g.}, $L_1\approx205\,$nm for Ag on Si). The confinement in the vertical direction is characterized by a symmetric exponential decay of the associated electric field intensity away from the film, extending a distance $\sim\lamp/4\pi$ regardless of the choice of materials and metal thickness (Fig.\ \ref{Fig1}a). The comparatively small number of electrons that support the plasmons in atomically-thin films makes them more susceptible to the environment, so that electrical gating with attainable carrier densities can produce significant plasmon shifts in single-atom-layer noble metals \cite{paper236}, while the addition and electrical gating of a graphene film results in dramatic modulation for thicker films up to a few nanometers \cite{paper277}. Likewise, the presence of an analyte can shift the plasmon resonance and introduce molecule spectral fingerprints enhanced by the near field of the plasmons, similar to what has been observed with graphene \cite{paper256}. However, besides such plasmon shifts, all of these applications require spectrally narrow plasmons, so that spectral modulation results in strong changes in light transmission or scattering, and this in turn demands the fabrication of high-quality films.

% formulas for modulation and sensing

%------------------------------------------------------ 
\section{RESULTS AND DISCUSSION}

\begin{figure*}
\includegraphics[width=1\textwidth]{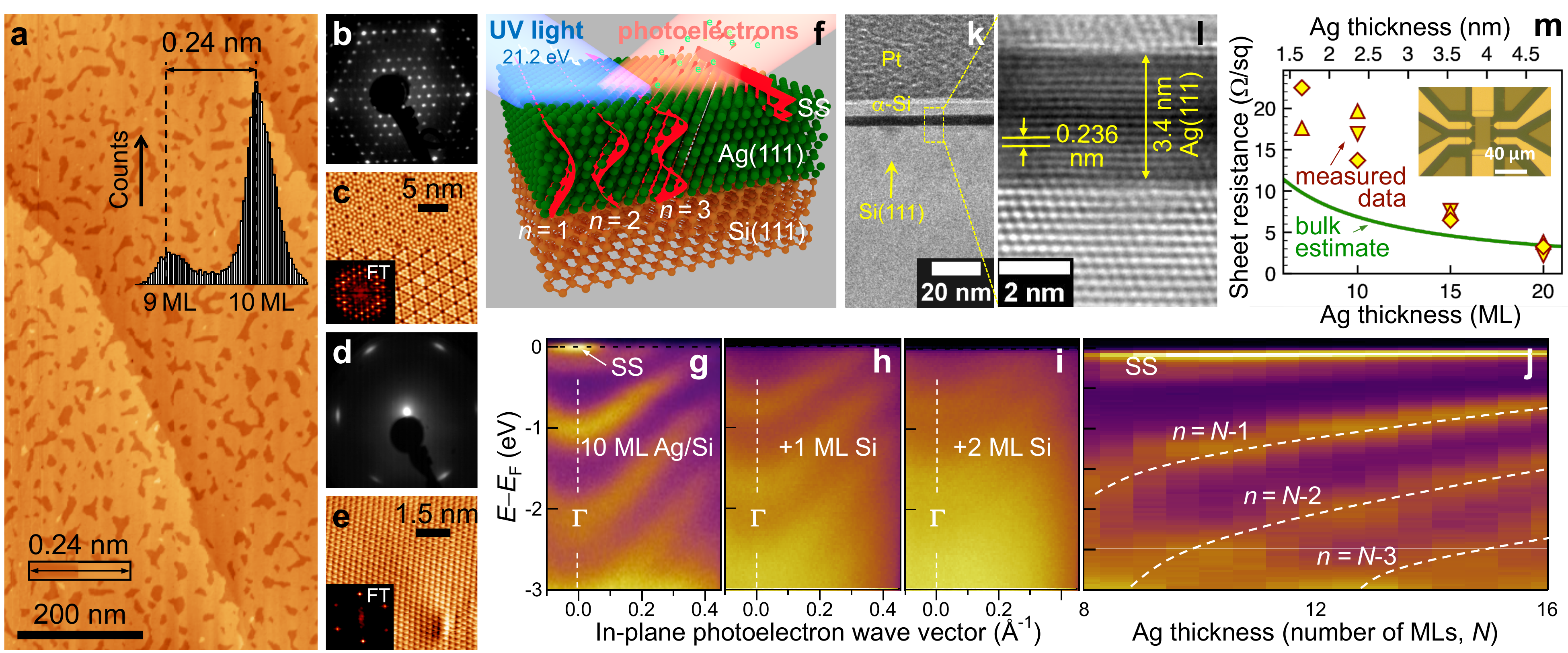}
\caption{{\bf Fabrication and characterization of atomically-thin crystalline silver films.} {\bf (a)} Scanning tunneling microscopy (STM) image of 10\,ML Ag(111) on Si. The histogram (upper inset) reveals a near completion of the 10$^{\rm th}$ layer (95\% area) with a small presence of 9\,ML (4\%, darker features) and 11\,ML (1\%, brighter features) islands (see color scale for out-of-plane distance). {\bf (b-d)} Low-energy electron diffraction (LEED) of (b) the bare Si(111) (7$\times$7 reconstruction) substrate and (d) after deposition of 10\,ML Ag(111), along with atomic-scale STM details for both surfaces (c and e, respectively). Fourier transforms of the STM images are shown in the lower-left corners.  {\bf (f)} Sketch of a Ag(111) film deposited on Si(111), along with its surface-state (SS) and the three lowest vertical quantum-well-state ($n=$1-3) wave functions, here probed through angle-resolved photoemission spectroscopy (ARPES). {\bf (g-i)} ARPES intensity as a function of electron energy relative to the Fermi energy (vertical scale) and parallel wave vector for (g) a 10\,ML Ag (111)/Si sample, and (h,i) after coverage with 1 and 2 MLs of Si. {\bf (j)} Evolution of the normal-emission ARPES intensity as a function of silver film thickness for 8-16\,ML Ag(111) on Si. Dashed curves are guides to the eye, corresponding the the top three states, with the state index $n$ (see f) varying with the number of layers $N$ as indicated by labels.. {\bf (k,l)} High-resolution transmission electron microscopy (HRTEM) images of the transversal cross section of a 14\,ML Ag(111)/Si sample, showing the silver atomic planes and their 0.236\,nm separation. {\bf (m)} Measured room-temperature sheet resistance for silver films consisting of $N=7$-20\,ML Ag(111)/Si (symbols), compared with the $293^{\circ}$C bulk estimate $\approx(68.7/N)\,\Omega/$sq (solid curve). Three different devices have been measured for each value of $N$, one of them is shown in the micrograph inset.} \label{Fig2}
\end{figure*}

We epitaxially grow high-quality crystalline Ag(111) films on a Si(111)-oriented wafer substrate \cite{NH97} with a controlled number of atomic monolayers under ultrahigh vacuum (UHV) conditions (see Methods). Through fine tuning of the growth parameters, we achieve films consisting of a single crystal domain on a cm$^2$ chip scale, as revealed by STM with atomic resolution (see Figs.\ \ref{Fig2}a and \ref{FigSI2}). The original $7\times7$ reconstruction of atomically flat Si (Fig.\ \ref{Fig2}c, where the upper and lower halves are empty and filled state images acquired with bias voltages of +2 V and -2 V, respectively) is removed upon Ag deposition, leaving an atomically flat Ag surface (Fig.\ \ref{Fig2}e) that preserves crystal lattice orientation ({\it cf.} Figs.\ \ref{Fig2}b and \ref{Fig2}d). We approach the targeted number of Ag(111) monolayers (10\,ML in Fig.\ \ref{Fig2}a) with just a $\sim5$\% fraction of regions differing by 1\,ML thickness. We complete structural characterization by imaging a cross section of the film using HRTEM, which reveals a preservation of defect-free ordering of atomic Ag(111) monolayers (Fig.\ \ref{Fig2}k,l) on the Si crystal substrate.

As plasmons are sustained by conduction electrons, we study the electronic band structure of the films, the small thickness of which produces discretization into a characteristic set of standing waves, encompassing vertical QWs \cite{C00_2,STM06,SAS14} (labeled by $n=1,\dots$ in the sketch of Fig.\ \ref{Fig2}f) and a surface-bound state (SS). Each of these QWs defines a band with nearly-free parabolic dispersion (effective mass $\approx1$), as revealed by ARPES (Fig.\ \ref{Fig2}g), which also show narrow lineshapes. We passivate our films with Si ($1.5$\,ML nominal thickness) in order to protect them during handling and patterning using e-beam nanolithography (see below). We note that  high-quality unpassivated Ag(111) films are stable during hours when brought from UHV to ambient conditions without patterning \cite{BBL93,SZK14}; however, strain in the Ag/Si interface eventually leads to film dewetting (within days), initiated by pinholes \cite{SZK14} and leading to silver oxides and formation of rough films. This protective Si layer is rapidly oxidized upon exposure to air, while the underlying Ag film is unaffected for weeks (see Fig.\ \ref{FigSI1}). We remark that the addition of the thin Si capping layer causes the SS to disappear but does not affect the QW states (Fig.\ \ref{Fig2}h,i). Control over thickness and high-quality of the films further allows us to experimentally observe a $\sim1/d$ scaling of the QW binding energies with increasing film thickness $d$ (Fig.\ \ref{Fig2}j), typical of a 1D particle-in-a-box system. We resolve QWs in all samples used in the present study, yielding an unambiguous determination of the number of layers in each Ag film. The presence and quality of the Ag film in the samples is further corroborated by ellipsometry measurements compared with bare Si substrates (see Fig.\ \ref{FigSI4}).

It is widely acknowledged fact that ultrathin metal films must experience strong surface scattering, and therefore see their electrical resistance sharply increased, as previous studies have indicated \cite{N1970,LHK94,BHG08,HGD10,DPS12}. In contrast, the high crystal quality of our films produces very low levels of the sheet resistance (Fig.\ \ref{Fig2}m), as revealed by four-probe measurements (see Fig.\ \ref{FigSI3}). In particular, we find the resistance to be only a factor of $\sim2$ higher than the estimate based on the bulk resistivity of silver for films as thin as 7\,ML Ag(111) ($1.65\,$nm thickness). Because the film quality does not open new channels for inelastic collisions compared with the bulk, we thus attribute this factor of 2 to defects introduced by the capping Si layer, which has reduced crystallinity (see top of Fig.\ \ref{Fig2}l), although some film damage during device fabrication cannot be ruled out. We attribute the large reduction of resistance in our films compared with previous studies, where films had a polycrystalline morphology, to the high crystallinity and absence of grain boundaries obtained by our followed epitaxial procedure (see Methods). The present results thus establish a much lower bound for the role played by surface scattering in the electrical resistance of high-quality crystalline silver films.

\begin{figure*}
\includegraphics[width=0.8\textwidth]{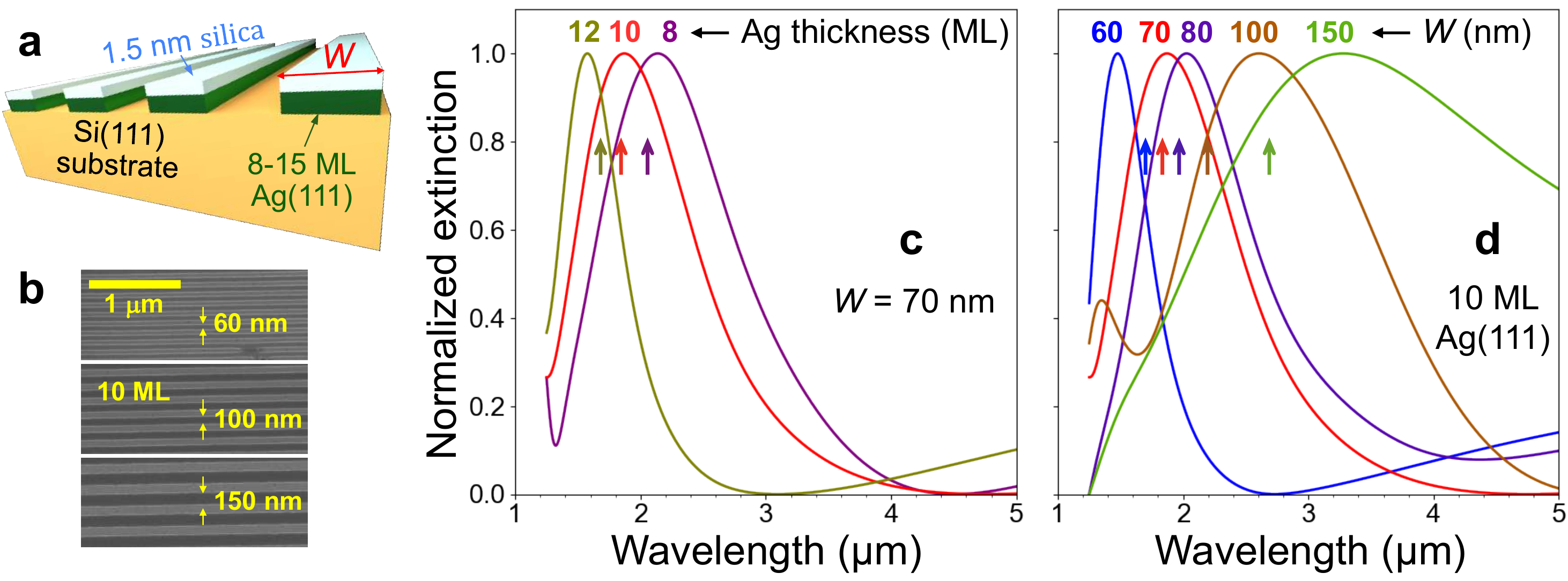}
\caption{{\bf Plasmons in atomically-thin crystalline silver nanoribbons.} {\bf (a)} Sketch of the ribbon arrays fabricated for this study. {\bf (b)} Scanning electron microscopy (SEM) images of some of the structures in a 10\,ML Ag(111)/Si sample, with the targeted ribbon width (dark areas) indicated in each case. {\bf (c)} Normalized optical extinction as experimentally measured for fixed ribbon width $W=70\,$nm and various metal film thicknesses (see labels). {\bf (d)} Same as (c) for fixed metal thickness (10\,ML) and varying ribbon width. Vertical arrows in (c,d) correspond to the analytical prediction of Eq.\ (\ref{l0W}) for the color-coordinated plasmon wavelengths.} \label{Fig3}
\end{figure*}

Plasmons in atomically-thin films are confined excitations with lateral wave vector $2\pi/\lamp$ greatly exceeding the light wave vector $2\pi/\lambda_0$, which prevents direct light-plasmon coupling. An additional source of lateral momentum is needed to break this optical momentum mismatch, such as that provided by a pattern in the films. In this work, we use e-beam nanolithography (see Methods) to carve ribbons with the desired range of widths $W\sim$50-500\,nm, which allow us to explore plasmon wavelengths $\lamp\approx2.7\,W$ (Fig.\ \ref{Fig1}b and Methods). The structure under consideration is sketched in Fig.\ \ref{Fig3}a, while scanning electron microscopy (SEM) images of some of the actual structures are shown in Fig.\ \ref{Fig3}b. The resulting measured optical spectra for different film thicknesses and ribbon widths are presented in Fig.\ \ref{Fig3}c,d, where plasmon redshifts are clearly observed when reducing the thickness or increasing the width, in qualitative agreement with the analytical formula
\begin{align}
\lambda_0\approx L_1\sqrt{2.7\,W/d}
\label{l0W}
\end{align}
(vertical arrows in Fig.\ \ref{Fig3}c,d), which predicts the light wavelength associated with the plasmon to scale linearly with the square root of the width-to-thickness aspect ratio $W/d$. This expression, which follows from the Drude model combined with the relation between $\lamp$ and $W$ (see Methods) \cite{BS17}, is in excellent agreement with a quantum-mechanical description of few-layer Ag films based upon the random-phase approximation combined with a realistic description of QWs in the films (see Fig.\ \ref{FigSI10}). Analytically calculated spectra (see Methods) have a similar level of agreement with measurements and nearly coincide with full electromagnetic simulations (see Fig.s\ \ref{FigSI8} and \ref{FigSI9}). Additionally, the spectra of Fig.\ \ref{Fig3}c,d reveal an increase in plasmon broadening with increasing ribbon width (see below).

\begin{figure*}
\includegraphics[width=1\textwidth]{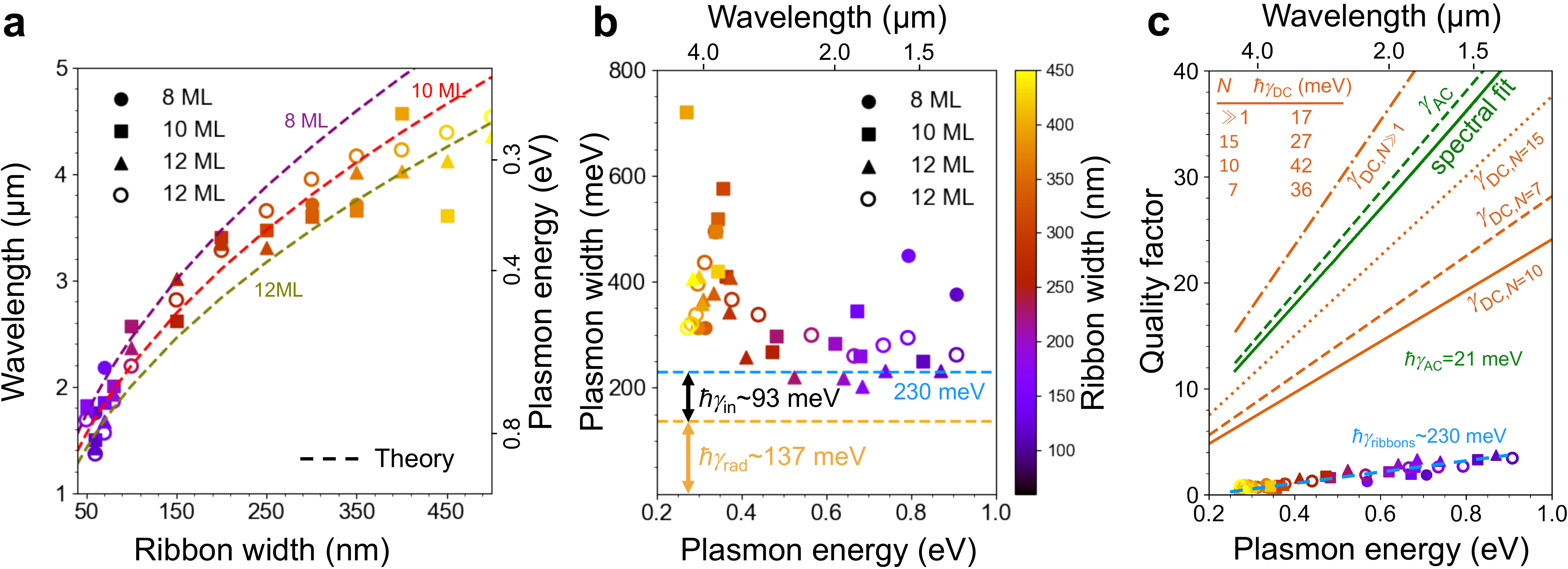}
\caption{{\bf Engineering the wavelength and quality factor $Q$ of plasmons in atomically-thin crystalline silver films.} {\bf (a)} Plasmon energy (right scale) and corresponding light wavelength (left scale) as a function of ribbon width. Experiment (symbols) is compared with simple analytical theory (dashed curves, Eq.\ (\ref{l0W})). {\bf (b)} Plasmon spectral width $\hbar\gamma$ as a function of plasmon energy $\hbar\omega$. Measured data from samples with various silver thicknesses are plotted using different symbols, with the ribbon width encoded in the color scale. The upper dashed horizontal line is a guide to the eye. The lower dashed line shows the radiative contribution to the damping $\hbar\gamma_{\rm rad}$ according to Eq.\ (\ref{gammarad}) for 10\,ML. {\bf (c)} Quality factor (peak energy divided by FWHM spectral width, $Q=\omega/\gamma$) as predicted by the Drude model for silver films using different input values of the damping rate $\gamma$ (see main text) compared with the experimental data taken from (b).} \label{Fig4}
\end{figure*}

These plasmon characteristics are consistently corroborated upon inspection of different samples (see Figs.\ \ref{FigSI5} and \ref{FigSI6}), the wavelengths and spectral widths of which are summarized in Fig.\ \ref{Fig4}. In particular, when plotting the observed plasmon wavelength as a function of ribbon width, we obtain a good agreement with Eq.\ (\ref{l0W}), despite deviations in individual structures, which we attribute to a variability in the actual width of the patterned ribbons. Additionally, we find a roughly constant plasmon width $\hbar\gamma\approx230\,$meV, which gives rise to a linear increase in $Q=\omega/\gamma$ with plasmon energy $\hbar\omega$ (see Fig.\ \ref{Fig4}b,c and quality-factor extraction procedure in Fig.\ \ref{FigSI7}). In our experiments, we find values of $Q$ approaching 4 at plasmon energies near 1\,eV. These spectrally narrow plasmons are made possible by the small thickness of our metal films combined with their crystalline quality. Indeed, polycrystalline films in the few nanometer range hardly reach $Q\sim1$ \cite{paper326}. Additionally, spatially confined NIR plasmons in noble metals require the use of high aspect ratios $W/d\sim20$; before the present study, high-quality structures could only be obtained for much thicker films, therefore involving larger $W$, and in consequence producing substantially broader plasmons due to coupling to radiation. Likewise, NIR plasmons in metallic colloids demand large particle aspect ratios, for which the observed quality factors are significantly smaller than 4 due to radiative losses as well \cite{paper300}, while in contrast to our films, the strategy of bringing the particle size to the few-nanometer range introduces additional plasmon quenching originating in finite-size effects \cite{SKD12} and thereby limiting the achievable $Q$.

The plasmon quality factors in our ribbon arrays are partially limited by radiative losses (see Fig.\ \ref{Fig1}c). Indeed, as shown in Methods, the total plasmon damping rate $\gamma=\gamma_{\rm in}+\gamma_{\rm rad}$ is the sum of an intrinsic component $\gamma_{\rm in}$ and a geometry-dependent radiative component $\gamma_{\rm rad}=\Gamma\times Wd/a$, where $\hbar\Gamma\approx88\,$meV/nm for Ag(111) films on silicon. For our experiments we fabricate ribbons with a period-to-width ratio $a/W=1.5$, which yields $\hbar\gamma_{\rm rad}\sim137\,$meV for 10\,ML films. This value is shown in Fig.\ \ref{Fig4}b as a lower dashed line; so we are left with an intrinsic damping $\hbar\gamma_{\rm in}\sim93\,$meV, which is still over 4 times larger than the bulk value of $21\,$meV derived from the measured permittivity of bulk silver \cite{JC1972}. We note that radiative losses should be negligible for arrays of large period-to-width ratio, thus suggesting a direct way to improve the quality factor with the same film quality (see Fig.\ \ref{FigSI8}).

The excess of intrinsic damping is presumably originating in sample damage incurred during the etching processes used for e-beam nanolithography (see Methods). Now, the question arises, how high can $Q$ be for confined plasmons based upon high-quality metal films consisting of a few atomic layers ({\it e.g.},  $<2$\,nm for 8\,ML Ag(111))? We address this question by comparing our measured $Q$'s with different estimates based upon the Drude expression $Q=\omega/\gamma_{\rm in}$ neglecting radiative losses (Fig.\ \ref{Fig4}c). Assuming the value $\hbar\gamma_{\rm AC}=21\,$meV obtained by fitting the measured Ag permittivity \cite{JC1972} to a Drude tail in the sub-1\,eV spectral region, we find $Q>40$ (an order of magnitude larger than those observed here), in agreement with predictions based on the estimate of $Q$ given by $-{\rm Im}\{\epsilon\}/{\rm Re}\{\epsilon\}$ \cite{MJK15}. This is also in good agreement with both the quality factors of spectra calculated in the long-wavelength limit (spectral fit) and the estimate obtained form the measured bulk DC conductivity ($\hbar\gamma_{{\rm DC},N\gg1}\approx17\,$meV). We note that the sheet resistance measured from our films (Fig.\ \ref{Fig2}m) leads according to the Drude model to predicted values $Q>20$ for 10\,ML Ag(111) films (see Methods) in the absence of radiative losses.

%------------------------------------------------------ 
\section{CONCLUDING REMARKS}

In brief, we report well-defined plasmons in atomically flat Ag(111) films grown on Si(111), with thickness as low as 8 ML ($\sim1.9\,$nm). The measured quality factors reach values $\sim4$. Further improvement of these results should include the exploration of thinner films down to 1-3 ML, which are however challenging because of the strain associated with the Ag/Si(111) interface. Following a two-step process ({\it i.e.}, deposition at low temperature $\sim100\,$K followed by annealing to 300\,K), we find the lowest thickness needed to produce atomically flat films using this procedure to be 6 ML. Nevertheless, 2ML Ag(111) films have been recently reported by employing a Ga/Si buffer layer \cite{HJQ14}, showing rather flat surfaces and well-defined quantum-well states \cite{SZJ18}. In our films, the crystalline quality of the fabricated Ag(111) films, which exhibit a clean electronic band structure consisting of quantized QWs, combined with the crystalline quality of the substrate, permit ruling out inelastic electron- and plasmon-scattering due to imperfections. However, the non-crystalline protecting capping layer can introduce inelastic coupling channels. Additionally, the etching processes used during e-beam lithography can cause sample damage, to which we attribute the reduction by half in film electrical conductance and by a factor of 5 in optical quality factor with respect to the maximum estimate in the studied spectral range, while another factor of $>2$ in quality factor can be gained by reducing radiative losses (e.g., by increasing the period-to-width ratio of the ribbon arrays). Further improvement in nanofabrication could therefore increase the achieved plasmon quality factors. Nevertheless, the plasmons here observed should be already sufficiently narrow to produce large electro-optical modulation in the NIR \cite{paper277}, while their reduced vertical and lateral size (down to $\sim20\,$nm and $\sim50\,$nm at 0.8\,eV, see Figs.\ \ref{Fig1}a and \ref{Fig4}a) are ideally suited for enhancing the interaction with neighboring molecules, thus holding great potential for optical sensing.

\section*{METHODS}

\noindent {\bf Fabrication of Atomically-Thin Silver Films.} Our Ag/Si(111) samples were prepared inside an UHV chamber at $1.0\times10^{-10}\,$mbar base pressure. We used $4\times12$\,mm$^2$ n-doped Si(111) chips with specific resistance 120-340\,$\Omega\,$cm as bare substrates. The dopant concentration of Si (1.3-$3.7\times10^{13}$\,cm$^{-3}$) was chosen to guarantee the electrical conduction required by surface science techniques, while not influencing the plasmonic performance of the silver films. Once inside the UHV chamber, the Si(111) chips were degassed overnight at 900\,K and subsequently flashed to 1400\,K for 20-30\,sec to remove the native silicon oxide. The sample temperature was slowly reduced to 600\,K, maintained at this temperature for 30\,min, and then cooled down to room temperature. This resulted in the formation of a defect-free, atomically clean Si(111) surface with a 7$\times$7 reconstruction. Silver atoms were sublimated from an electron-bombardment evaporator, which was calibrated to sub-monolayer accuracy using a quartz microbalance monitor in combination with probing the distinct 1-2\,ML Ag/Cu(111) surface states by photoemission \cite{SCV05}. Silver films were grown on Si(111) following this two-step process. The Si(111) substrate was kept at 100-120\,K during Ag deposition and slowly annealed to room temperature afterwards \cite{MH08_3}. The deposition rate was $\sim0.3\,$ML/min, although a similar film quality was obtained within the 0.1-0.5\,ML/min range; the crucial parameter here is the deposition temperature, which was required to be $\sim100\,$K.

\begin{figure*}
\includegraphics[width=0.7\textwidth]{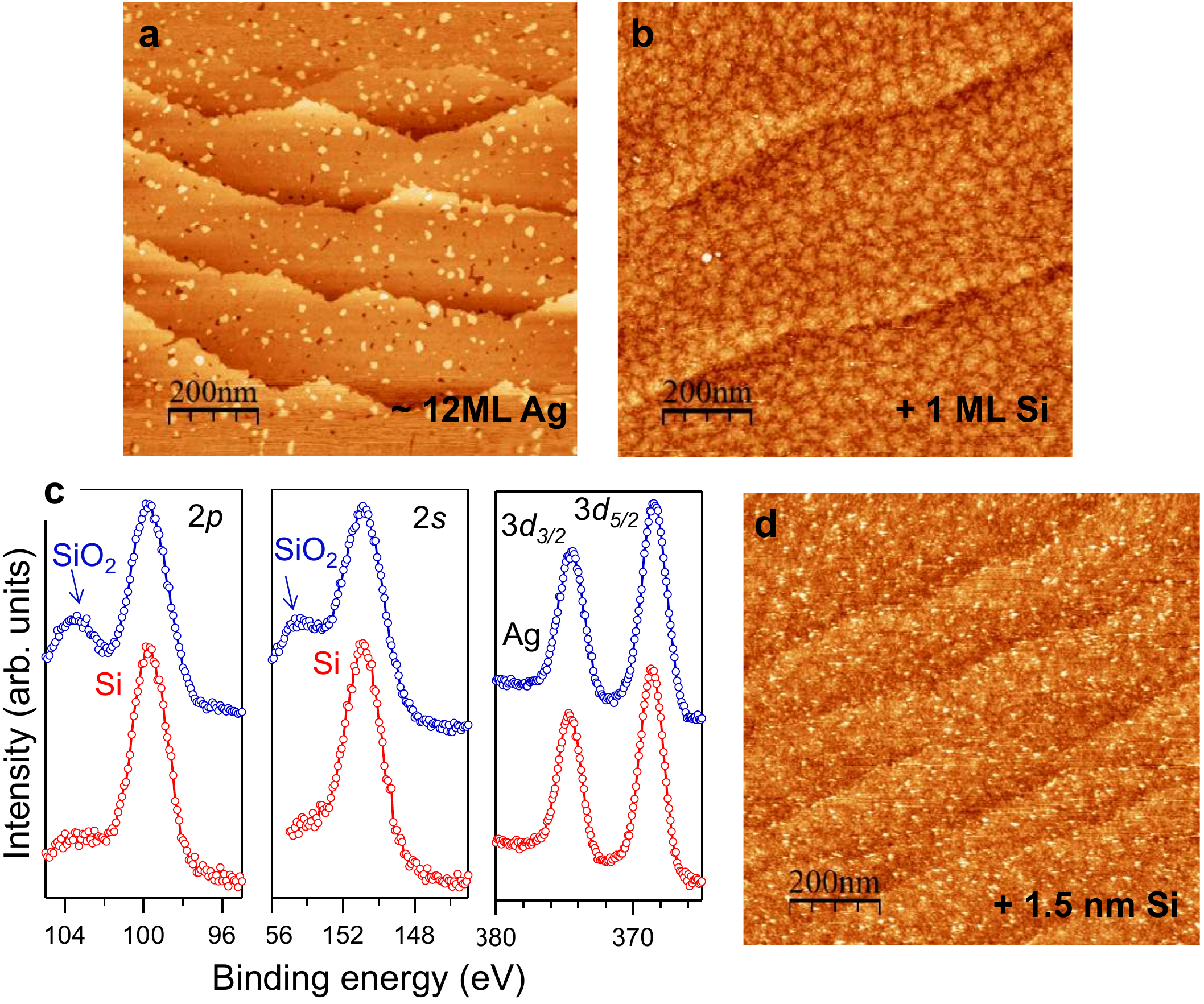}
\caption{{\bf Surface quality upon Si capping and after exposure to ambient conditions.} {\bf (a,b)} Large-scale ($1\times1\,\mu$m$^2$) STM images of a 12\,ML Ag(111) film (a) before and (b) after capping with $\sim1$\,ML Si. Atomic Si patches are shown to cover  the surface. {\bf (c)}  X-ray photoemission spectra (XPS) for the sample shown in (b), collected after 20\,min (red) and 15 days (blue) exposure to ambient atmosphere conditions. A SiO$_{2}$ peak is clearly developed, while Ag d-levels are not affected. {\bf (d)}  STM image of the sample after capping with $\sim1.5$\,nm Si.}
\label{FigSI1}
\end{figure*}

\begin{figure*}
\includegraphics[width=0.75\textwidth]{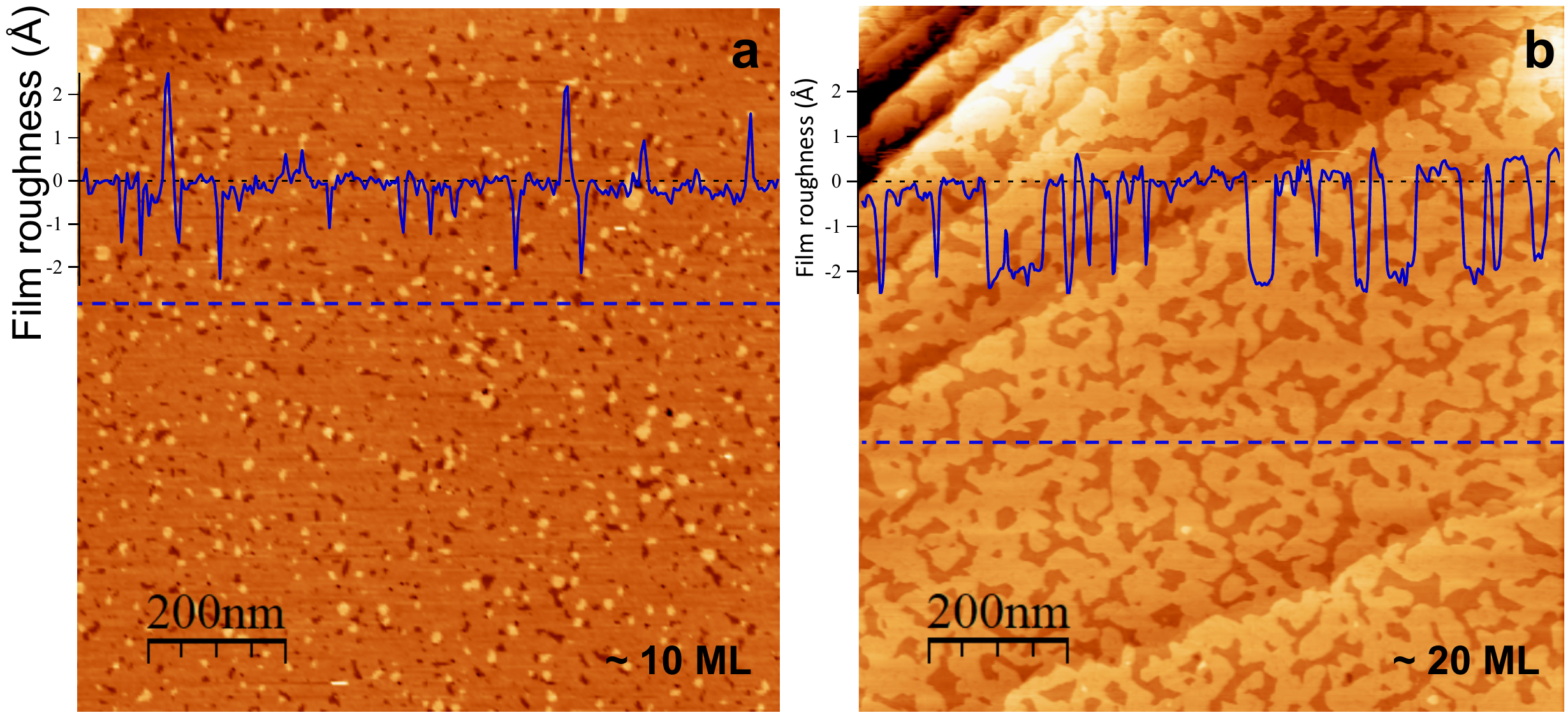}
\caption{{\bf Additional examples of characteristic surface thickness distributions.} Large-scale ($1\times1\,\mu$m$^2$) STM images of Ag(111) films with nominal thickness of (a) 10\,ML and (b) 20\,ML. The  profiles taken along the blue-dashed lines (superimposed blue curves) reveal a maximum variation of film thickness $\pm0.24\,$nm, corresponding to $\pm1\,$ML.}
\label{FigSI2}
\end{figure*}

\noindent {\bf Surface-Science Characterization.} The atomic and electronic structure of the Si substrate and the grown Ag films were characterized by LEED, STM, and ARPES. STM data were collected using an Omicron VT setup operating at room temperature. ARPES measurements were performed using a SPECS Phoibos 150 electron analyzer equipped with a monochromatized He gas discharge lamp operating at the He I$\alpha$ excitation energy (21.2\,eV), with an electron energy and angular resolution of 30\,meV and $0.1^\circ$, respectively. The diameter of the UV light beam was $\sim0.5\,$mm at the sample surface. Sample transfer between STM and ARPES setups was made without breaking UHV conditions. Prior to atmosphere exposure, the samples were capped by a Si protection layer ($1.5\,$nm nominal thickness), evaporated by direct heating of a Si chip with the same doping level as the substrate. The robustness and aging of the films was monitored by X-ray photoemission spectroscopy (XPS, see Fig.\ \ref{FigSI1}).

\noindent {\bf HRTEM Characterization.} Electron-transparent ($<50\,$nm thickness) cross-sectional lamellas of the samples were prepared by first sputtering a platinum layer for protection, followed by carving using a FEI Helios NanoLab 600 dual beam SEM/focused-ion-beam (FIB) system. After transfer of the lamellas to a copper grid, they were imaged using a JEOL JEM-2100 HRTEM operated at 200\,kV.

\begin{figure*}
\includegraphics[width=0.35\textwidth]{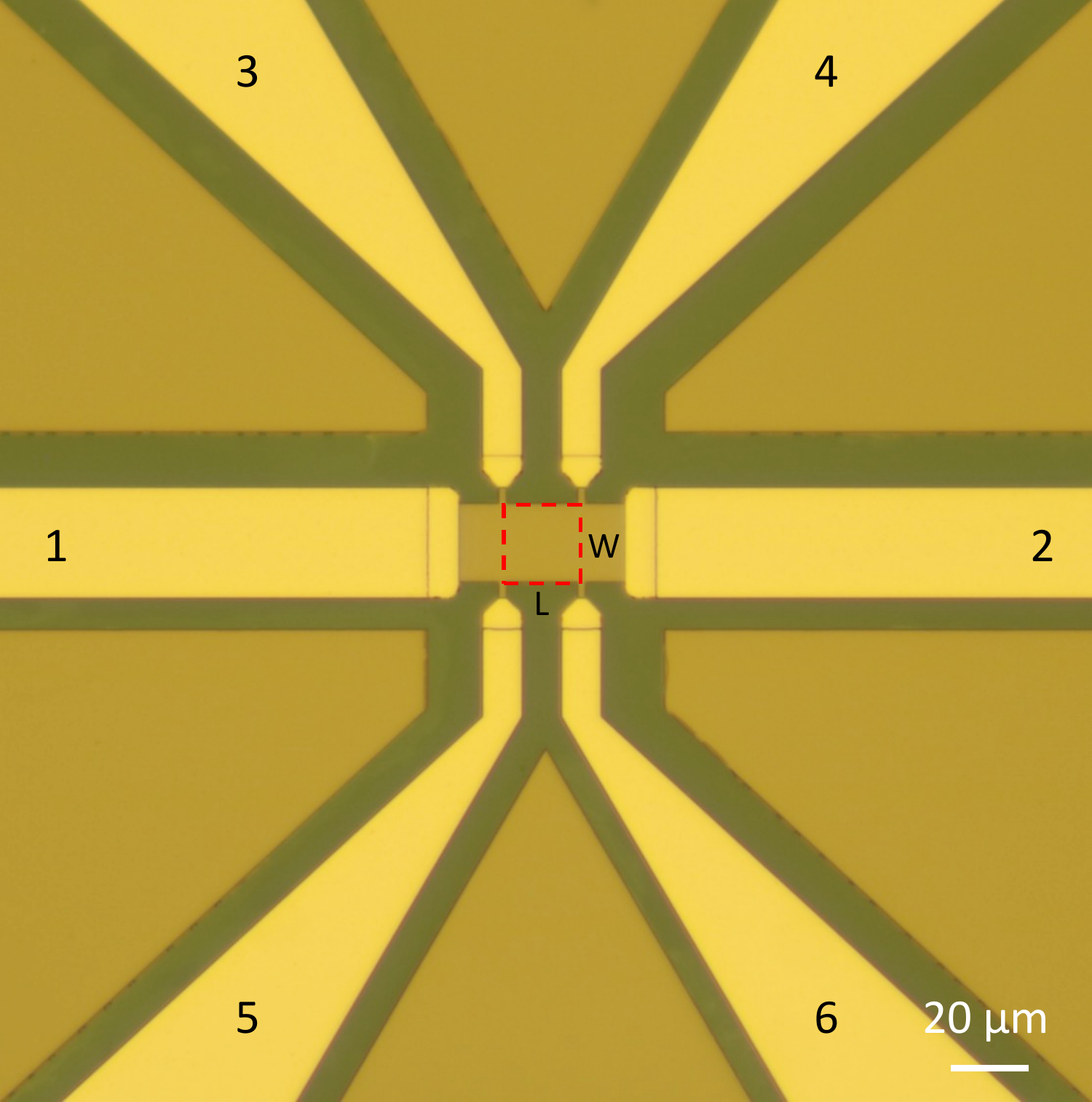}
\caption{{\bf Sheet resistance measurements}. This optical image shows the footprint of a characteristic four-probe Hall-bar structure device used for the measurement of the sheet resistance. The Hall-bar structure consists of 6 electrodes. Electrodes 3 to 6 are voltage probes. The resistance $R$ of the region highlighted with a red rectangle is calculated using the expression $R=V_{34}/I_{12}$, where $V_{34}$ is the voltage difference between probes 3 and 4, whereas $I_{12}$ denotes the current flowing through electrodes 1 and 2. The sheet resistance $R_{\rm s}$ of the ultrathin silver film is given by $R_{\rm s}=R W/L$, where $L=W=20\,\mu$m (see image), so that $R=R_{\rm s}$.}
\label{FigSI3}
\end{figure*}

\noindent {\bf Sheet Resistance Measurements.} Ultrathin silver films were etched into a Hall-bar structure by argon plasma using an Oxford Plasmalab 100 reaction-ion etching (RIE) system. A Poly (methyl methacrylate) (PMMA) layer was used as the etch mask. Contact electrodes were formed by depositing a Cr/Au/Al (3/60/190\,nm) layer followed by lift-off. All structures were patterned by a Raith EBPG 5000+ e-beam lithography system. A four-probe scheme \cite{S1958} (Fig.\ \ref{FigSI3}) was used to extract the sheet resistance. The electrical characterization was performed in a Lakeshore probe station operating at $7\times10^{-5}\,$mbar. An Agilent B1500A semiconductor parameter analyzer was used for all electrical measurements.

\noindent {\bf Electron-Beam Nanolithography.} Passivated silver-film chips were uniformly spin-coated with $\sim100\,$nm ZEP520A resist for 1 min at 6000 rpm. Ribbons were then written using a RAITH150-Two e-beam lithography system, followed by development in amyl acetate and reactive-ion etching for $\sim1$\,min with an Ar and CHF$_3$ mixture in a RIE Oxford Plasmalab 80 Plus system. Periodic arrays of 50-1000\,nm-wide ribbons were fabricated with a footpring of $200\times200\,\mu$m$^2$ per sample and $\sim1.5$ period-to-width ratio. Importantly, although standard procedures usually involve baking at 150-180$^\circ$C after spin-coating to induced a phase transition to glass in the resist, we skipped this step to avoid Ag film damage, at the expense of having a more fragile resist that required careful calibration of the RIE gas mixture and etching time to preserve the etching mask.

\begin{figure*}
\includegraphics[width=0.85\textwidth]{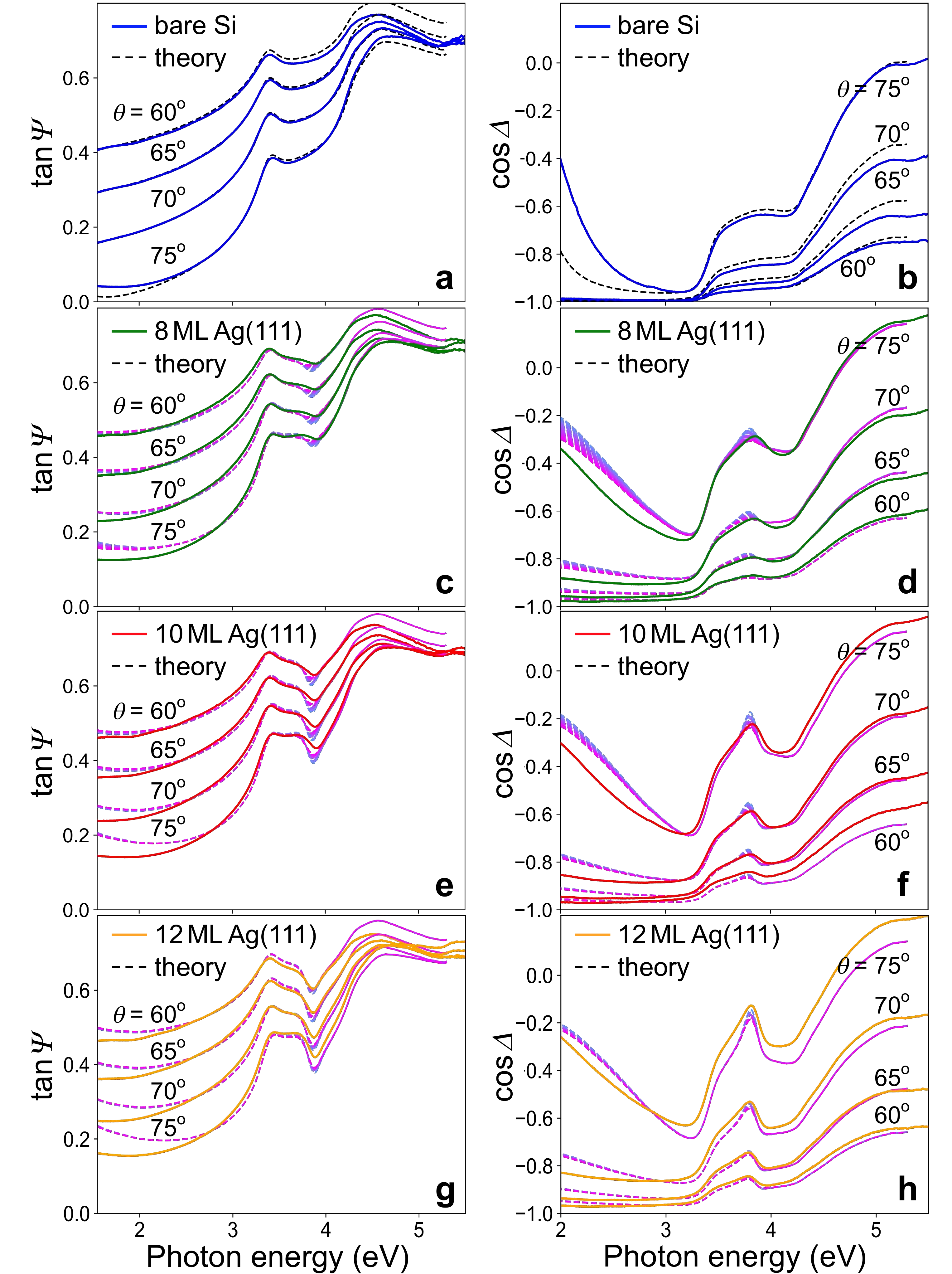}
\caption{{\bf Ellipsometry measurements.} We plot the measured ellipsometric parameters (a,c,e,g) $\tan\Psi$ and (b,d,f,h) $\cos\Delta$ for (a,b) bare Si and (c-h) Ag/Si samples containing 8-12\,ML Ag(111) (solid curves, see legends) compared with theory (dashed curves). These measurements are performed prior to any lithography process. The ellipsometry parameters are defined by the relation $r_{\rm p}/r_{\rm s}=\left(\tan\Psi\right)\,\exp(\ii\Delta)$, where $r_\sigma$ is the Fresnel reflection coefficient for $\sigma=$s, p polarization. Measurements are performed for each sample at different angles of incidence in the $\theta=60^\circ$-$75^\circ$ range (see labels). Simulations are carried out assuming a silica layer with a fixed thickness of 1.5\,nm capping the Ag films, which have in turn a thickness $N\times0.236\,$nm ({\it i.e.}, taking the bulk spacing between Ag atomic planes). The dielectric functions of these materials, as well as that of the crystalline Si substrate, are taken from the SOPRA ellipsometry data base \cite{SOPRA}, after adding a correction to the Ag permittivity intended to use a possible increase in Drude damping as a fitting parameter ({\it i.e.}, we replace the Ag permittivity $\epsilon_{\rm Ag}(\omega)$ by $\epsilon_{\rm Ag}(\omega)+\omega_{\rm bulk}^2/\omega(\omega+\ii\gamma)-\omega_{\rm bulk}^2/\omega(\omega+\ii\eta\gamma)$, where $\hbar\omega_{\rm bulk}=9.17\,$eV and $\hbar\gamma=0.021\,$eV are experimental Drude parameters \cite{JC1972}). Theoretical results (dashed curves) are presented for different ranges of $\eta$, with values increasing from light-blue to purple curves: $\eta=10$-25 in the 8\,ML film; $\eta=1$-10 in the 10\,ML film; and $\eta=1$-5 in the 12\,ML film.}
\label{FigSI4}
\end{figure*}

\begin{figure*}
\includegraphics[width=0.9\textwidth]{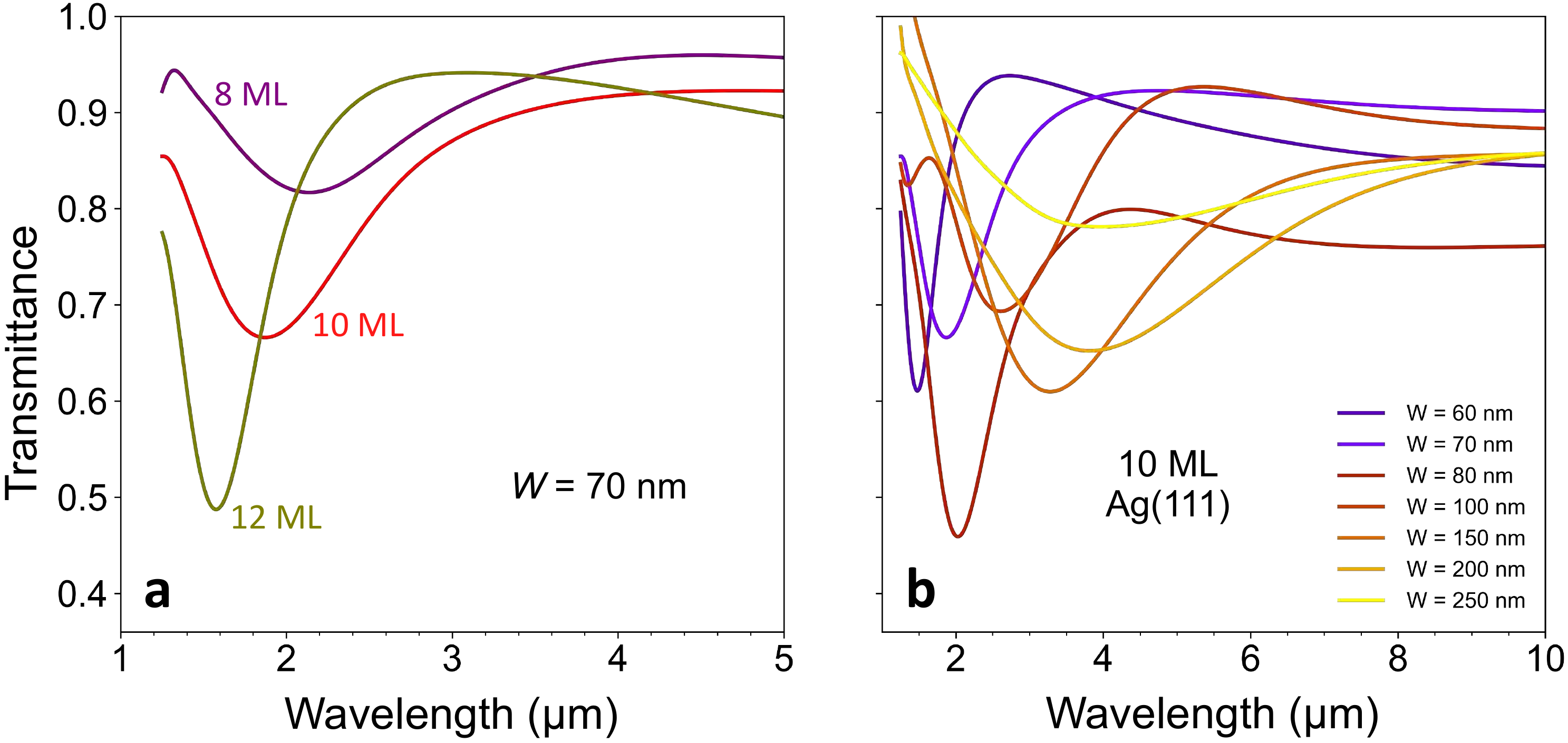}
\caption{Measured transmission spectra used in Fig.\ \ref{Fig3}c,d of the main text before normalization.}
\label{FigSI5}
\end{figure*}

\begin{figure*}
\includegraphics[width=0.9\textwidth]{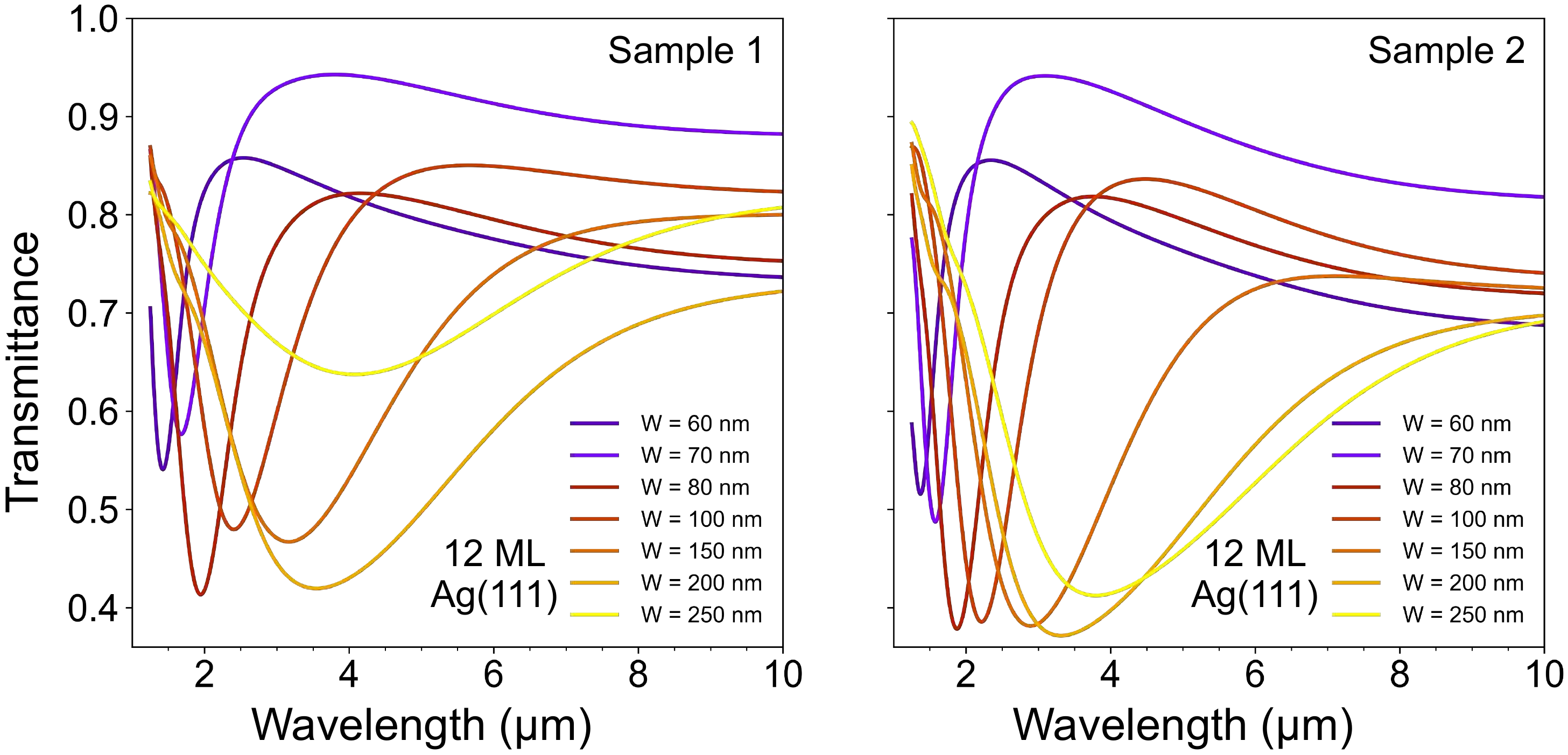}
\caption{Measured transmission spectra for 12\,ML Ag(111) on Si taken from two different samples (left and right panels, respectively) for the same ribbon widths (see legend).}
\label{FigSI6}
\end{figure*}

\begin{figure*}
\includegraphics[width=1\textwidth]{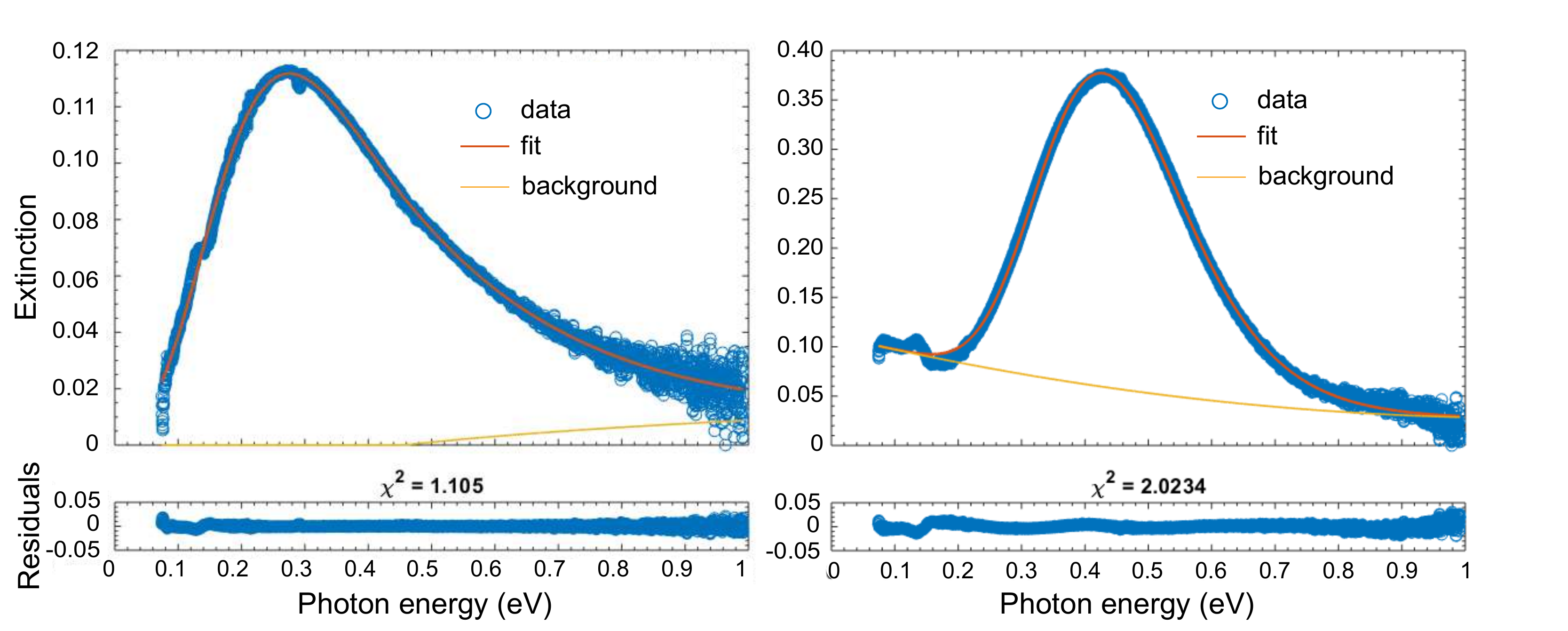}
\caption{{\bf Quality factor extraction}. We show two examples of quality factor extraction from experiment. Measured extinction data (blue circles) are fitted by an asymmetric line-shape function (red curves) after subtraction of a background (yellow curves). In more detail, we use a Skew-Pseudo-Voigt function for the extinction (a combination of Lorentzian and Gaussian profiles, $f_{\rm L}(\omega)=(\gamma_{\rm L}/\pi)/\left[(\omega-\omega_0)^2+\gamma_{\rm L}^2\right]$ and $f_{\rm G}(\omega)=\exp\left[-(\omega-\omega_0)^2/2\sigma_{\rm G}^2\right]/\left(\sqrt{2\pi}\sigma_{\rm G}\right)$, respectively), defined as $f(\omega)=(1-\eta)f_{\rm G}(\omega)+\eta f_{\rm L}(\omega)$, where $\sigma_{\rm G}=\gamma_{\rm L}/\sqrt{2\log2}$ and $\gamma_{\rm L}=\Gamma_0/\left[1+\ee^{-a(\omega-\omega_0)}\right]$. We use $0\leq\eta\leq1$, $a$, $\Gamma_0$, and $\omega_0$ as fitting parameters. Parabolic profiles are used to subtract a background, with parameters defined to match the end data points in the fitted spectral range ({\it i.e.}, the red curves are the sum of $f(\omega)$ and this background). The lower plots show the residuals (data minus fit) and standard deviation. The quality factors are determined as the ratios of peak energy to full-width at half-maximum (fwhm) in the energy spectra of the fitting curves $f(\omega)$.}
\label{FigSI7}
\end{figure*}

\noindent {\bf Optical Characterization.} We used a SOPRA GES-5E system to perform ellipsometry  (Fig.\ \ref{FigSI4}) for incidence angles in the 60-75$^\circ$ range over the UV-NIR photon energy region (1.5-5\,eV). Optical transmission/reflectance spectra (Figs.\ \ref{FigSI5} and \ref{FigSI6}) were collected using a BRUKER HYPERION fourier-transform infrared (FTIR) spectrometer operating in the 1.3-$17\,\mu$m range.

\noindent {\bf Analytical Simulations.} The plasmon dispersion relation (parallel wave vector $\kpar$ as a function of frequency $\omega$) of a homogeneous thin film is given in the quasistatic limit by \cite{paper235}
\[\kpar=\frac{\ii\omega(\epsilon_1+\epsilon_2)}{4\pi\sigma},\]
where $\epsilon_1$ and $\epsilon_2$ are the permittivities of the media on either side of the film, while $\sigma$ is the 2D conductivity. Assuming local response, we write the latter as
\[\sigma=(\ii\omega/4\pi)(1-\epsilon)d,\]
which is proportional to the film thickness $d$, and where $\epsilon$ stands for the metal permittivity; this is an excellent approximation for the materials and film thicknesses under consideration even when compared with quantum-mechanical simulations (see Fig.\ \ref{FigSI10}). Adopting the Drude model \cite{AM1976}, we approximate $\epsilon\approx1-\omega_{\rm bulk}^2/\omega(\omega+\ii\gamma_{\rm in})$ in terms of the bulk plasma frequency $\omega_{\rm bulk}$ and the intrinsic damping rate $\gamma_{\rm in}$ (assuming $\omega\ll\omega_{\rm bulk}$), which leads to the dispersion relation $\kpar d\approx(\epsilon_1+\epsilon_2)\,\omega(\omega+\ii\gamma_{\rm in})/\omega_{\rm bulk}^2$, and this in turn allows us to write the in-plane plasmon wavelength defined by $\lamp=2\pi/{\rm Re}\{\kpar\}$ as $\lamp=d\,(\lambda_0/L_1)^2$ ({\it i.e.}, Eq.\ (\ref{lpl0}) in the main text), where \[L_1=\sqrt{2\pi(\epsilon_1+\epsilon_2)}\;\frac{c}{\omega_{\rm bulk}}\] and $\lambda_0$ is the free-space light wavelength. For Ag films ($\hbar\omega_{\rm bulk}\approx9.17\,$eV \cite{JC1972}) deposited on silicon ($\epsilon_1\approx12$) and coated with ZEP502A resist ($\epsilon_2\approx2.4$), we find $L_1\approx205\,$nm, which renders $\lamp\ll\lambda_0$ at light wavelengths below $\sim5\,\mu$m when $d$ spans a few atomic layers (below $\sim15\,$ML), thereby justifying our using the quasistatic limit, although retardation effects can become apparent for longer wavelengths and thicker films. Incidentally, the resist is not removed from the samples before plasmon measurements, but the penetration depth $\lamp/4\pi$ is smaller than the resist thickness ($\sim$100\,nm), thus justifying the use of the resist permittivity in the above expression for $L_1$.

We remark that the above results assume a small film thickness $d$ compared with the plasmon wavelength $\lamp$, while the reduction of the metal film response to a surface conductivity is valid if $d$ is also small compared with the skin depth $\lambda_0/(2\pi{\rm Im}\{\sqrt{\epsilon}\})\approx c/\omega_{\rm bulk}\sim20\,$nm in Ag. Additionally, in the quasistatic limit, the electric field $\Eb$ is longitudinal ($\nabla\times\Eb=0$) and divergenceless ($\nabla\cdot\Eb=0$), therefore displaying a symmetric pattern relative to the negligibly-thick film (we refer to a recent study \cite{paper331} for more details). In particular, the electric field associated with the plasmon has symmetric (antisymmetric) in-plane (out-of-plane) components with respect to the normal coordinate $z$ and admits the expression \cite{paper331} $\propto\left[\xx+\ii\,{\rm sign}(z)\zz\right]\ee^{\kpar(\ii x-|z|)}$ for propagation along the in-plane direction $x$, from which an exponential decay away from the film is predicted with a $1/e$ fall in intensity at a distance $\lamp/4\pi$ from the film (see Fig.\ \ref{Fig1}a in the main text). We note that the field is however asymmetric if the film thickness is not small compared with both the plasmon wavelength and the metal skin depth. The above expression for the field also allows us to write the in-plane plasmon propagation distance (for $1/e$ decay in intensity) as $\Lp=1/2{\rm Im}\{\kpar\}$. Using the dispersion relation noted above, we find $\Lp=\lamp L_2/\lambda_0$, where $L_2=c/2\gamma_{\rm in}$ ({\it e.g.}, taking $\hbar\gamma_{\rm in}=21\,$meV for Ag, as obtained from optical data \cite{JC1972}, we have $L_2=4.7\,\mu$m); the propagation distance is then $L_2/\lambda_0$ (independent of metal thickness) times the plasmon wavelength (proportional to metal thickness). Incidentally, a plasmon lifetime $1/\gamma_{\rm in}$ is directly inherited from the Drude model in the absence of radiative losses (a good approximation for $\lamp\ll\lambda_0$) and substrate absorption (Si losses are negligible in the studied spectral range within the $\lamp/4\pi$ plasmon penetration depth), leading to a plasmon quality factor (frequency-to-width ratio) $Q=\omega/\gamma_{\rm in}$. This relation is used in Fig.\ \ref{Fig4}c of the main text with various estimates for $\gamma_{\rm in}$ (see below as wel). We also find useful to write the propagation distance as $\Lp=\lamp Q/4\pi$.

\begin{figure*}
\includegraphics[width=0.45\textwidth]{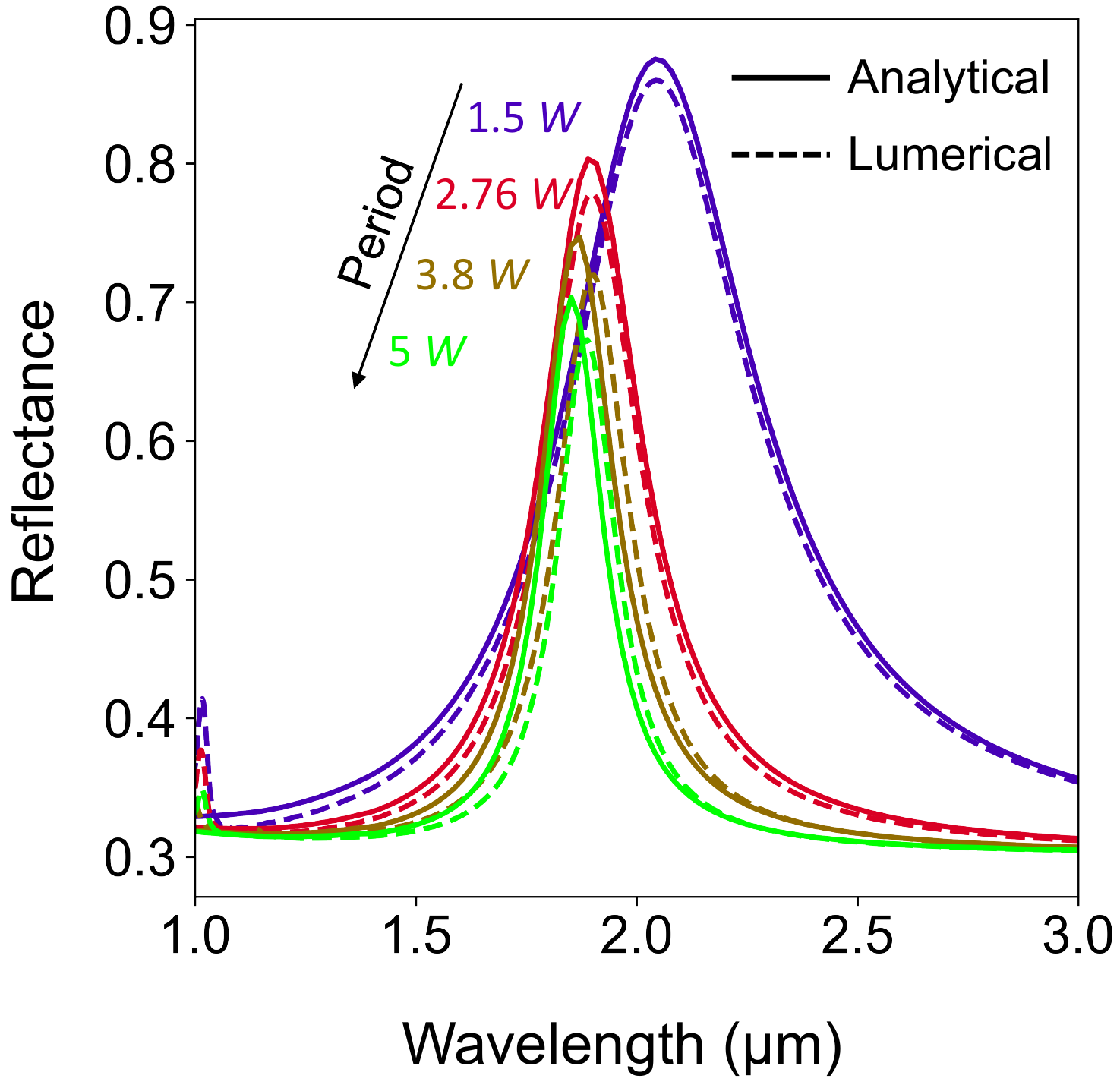}
\caption{{\bf Effect of ribbon separation.} Our experiments are performed for ribbon arrays with a period-to-width ratio $a/W\sim1.5$. The period of the array has a relatively mild influence on the position of the plasmon, but this parameter produces a substantial effect on the plasmon width, as observed in the simulations presented in this figure for ribbons of fixed width $W=70\,$nm and metal thickness $d=10$\,ML. (See caption of Fig.\ \ref{FigSI9} for more details.) As explained in the Methods section, the spectral width is the sum of an intrinsic damping rate $\gamma_{\rm in}$ (determined by the metal quality, down to $\hbar\gamma_{\rm in}\approx93\,$meV in this study, to be compared with $\approx21$\,meV in bulk silver \cite{JC1972}) plus a radiative damping rate (see Fig.\ \ref{Fig4} in the main text); the latter is estimated as $\gamma_{\rm rad}=\Gamma\times Wd/a$, where $\hbar\Gamma\approx88\,$meV/nm for Ag(111) films on silicon (see main text), yielding $\hbar\gamma_{\rm in}\approx137\,$meV in the present case (i.e., $d=10$\,ML and $a/W=1.5$). Increasing the period $a$ is thus a good strategy for increasing the plasmon quality factor.}
\label{FigSI8}
\end{figure*}

\begin{figure*}
\includegraphics[width=0.65\textwidth]{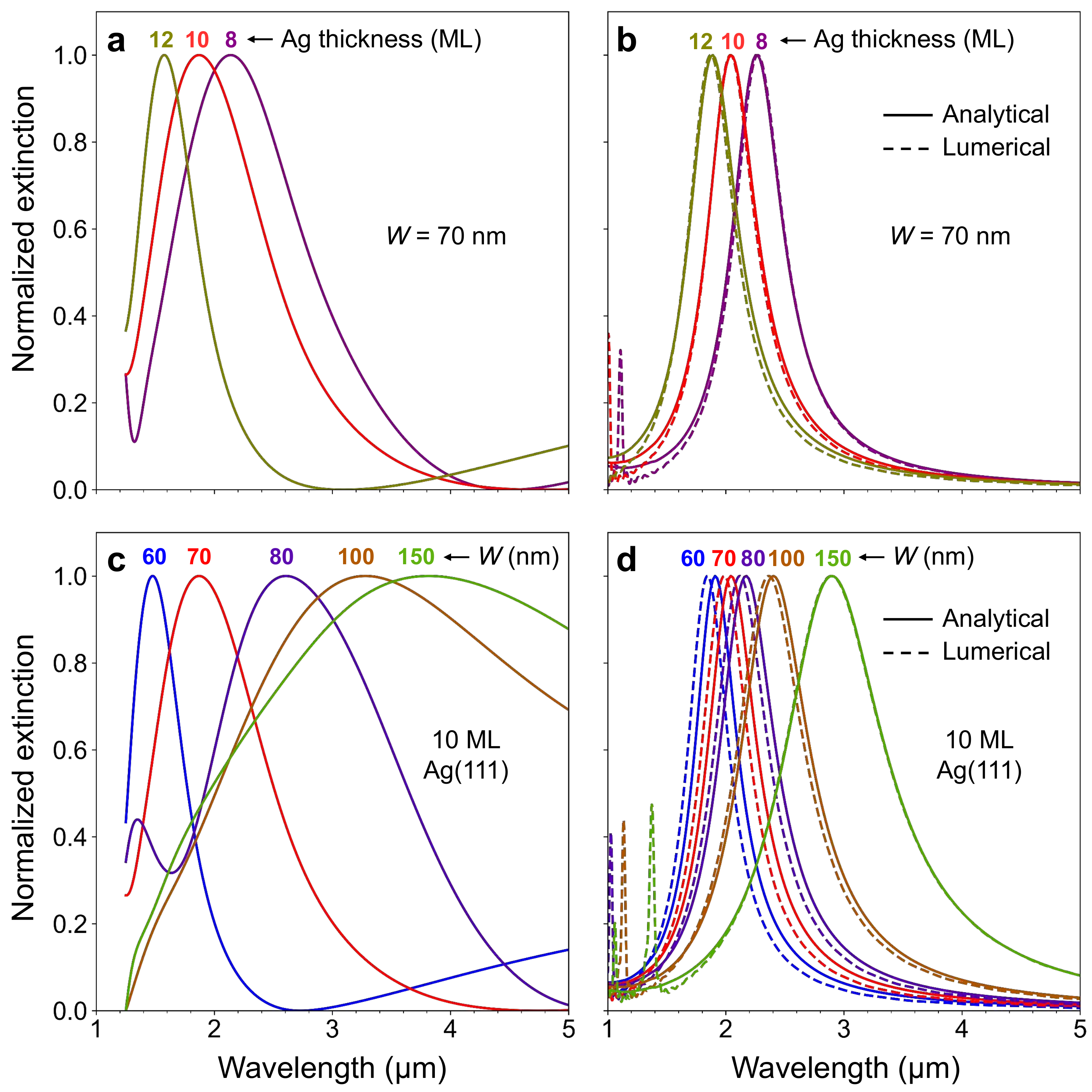}
\caption{{\bf Comparison with analytical and numerical simulations.} We compare measured spectra (a,c, directly reproduced from Fig.\ \ref{Fig3} in the main text) with theory (b,d). The latter shows the normalized extinction obtained with two different levels of theory: the analytical procedure detailed in the Methods section (solid curves) and numerical simulations carried out with the electromagnetic software Lumerical. We use tabulated optical data for the dielectric functions of silver \cite{JC1972} and crystalline silicon \cite{AS1983}. Calculations are performed for ribbon arrays with a period-to-width ratio of 1.5. Ribbons are assumed to be coated by a 1.5\,nm layer of $\epsilon=2$ dielectric.}
\label{FigSI9}
\end{figure*}

\begin{figure*}
\includegraphics[width=0.4\textwidth]{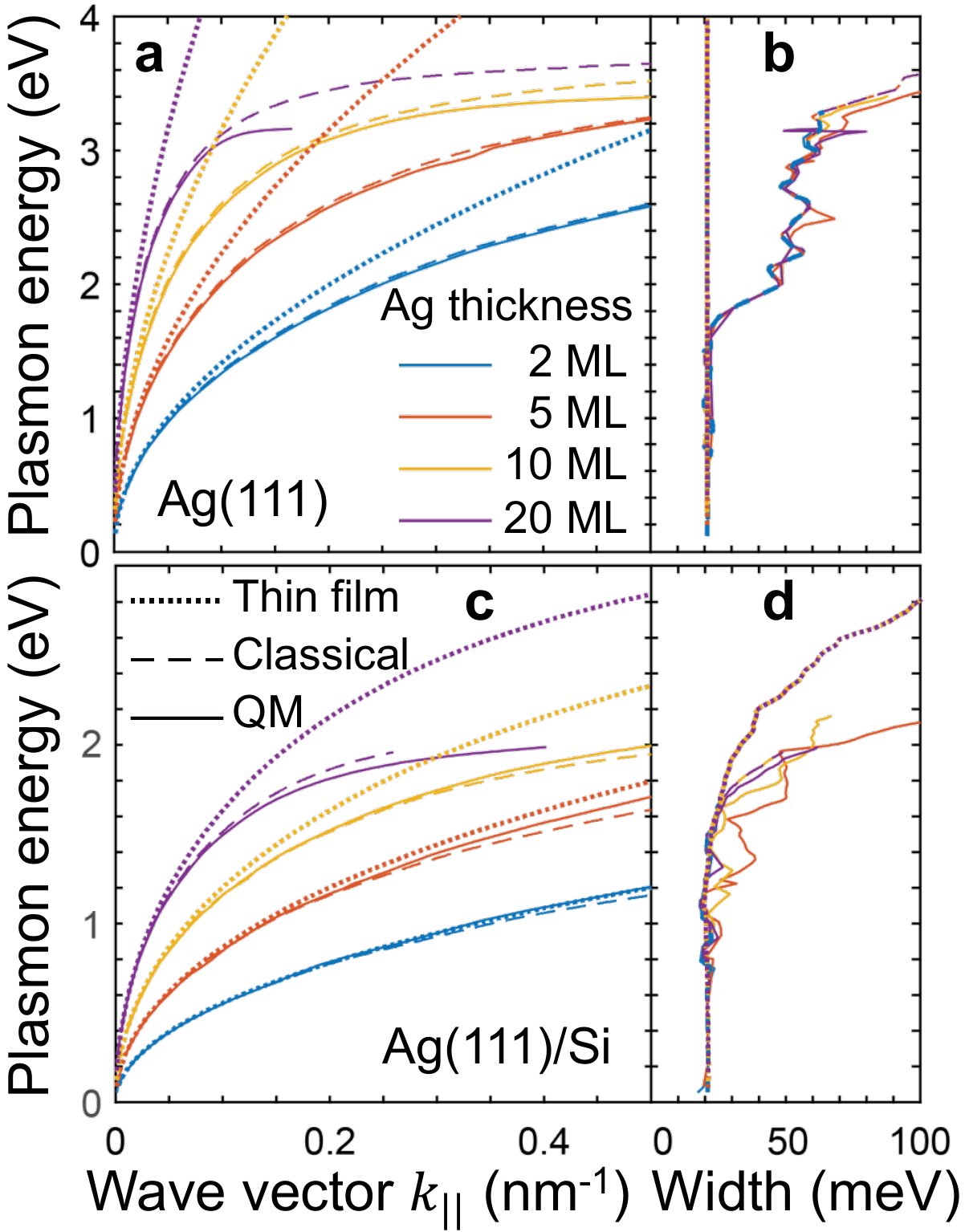}
\caption{{\bf Quantum mechanical (QM) vs classical description of plasmons in atomically-thin silver films.} We compare QM calculations (solid curves) with classical simulations (broken curves) for self-standing (a,b) and Si-supported (c,d) Ag films consisting of $N=2$-20\,ML Ag(111) (see legend). We plot the plasmon dispersion relation in (a,c) and the spectral width in (b,d). Details of the QM model are given elsewhere \cite{paper329}; in brief, we use the random-phase approximation (RPA) to describe the film response, introducing the conduction electron wave functions as input, and incorporating a background local dielectric response to account for Ag d-band polarization and the Si substrate; the wave functions are in turn calculated assuming in-plane parabolic dispersion, but capturing layer corrugation through a model potential along the out-plane direction that reproduces the most salient features of the electronic band structure in Ag surfaces \cite{CSE99}. The classical theory (dashed curves) assumes a local response of the film with thickness $d=N\times0.236\,$nm based upon the frequency-dependent Ag dielectric function $\epsilon_{\rm Ag}(\omega)$. We use tabulated optical data for the dielectric functions of silver \cite{JC1972} and crystalline silicon \cite{AS1983}. The thin-film dotted curves are calculated also classically by replacing the film by a zero-thickness layer of conductivity $\sigma=(\ii\omega d/4\pi)(1-\epsilon)$ with $\epsilon\approx1-\omega_{\rm bulk}^2/\omega(\omega+\ii\gamma)$  ({\it i.e.}, absorbing the entire dielectric response of the actual film into a 2D conductivity) for parameters $\hbar\omega_{\rm bulk}\approx9.17\,$eV and $\hbar\gamma=21\,$meV corresponding to Ag \cite{JC1972}; incidentally, we neglect d-band polarization in this expression ({\it i.e.}, the leading term of $1$ should be replaced by $\sim4$) because it does not affect the plasmon dispersion significantly below $\sim1.5$\,eV photon energy but allows us to obtain simpler analytical expressions.}
\label{FigSI10}
\end{figure*}

For ribbon arrays, plasmons are excited under transverse polarization ({\it i.e.}, with the electric field oriented across the width of the ribbons, see Fig.\ \ref{Fig1}b in the main text), whereas a featureless weak absorption is produced when the incident light field is parallel to the ribbons. Consequently, we concentrate on the former in what follows and adopt a previously reported model \cite{paper235} to calculate the normal-incidence transverse-polarization transmission coefficient as
\begin{align}
t=\frac{1}{\bar{n}}\left[1+\frac{\ii S}{\tilde{\alpha}^{-1}-G}\right],
\label{tarray}
\end{align}
where $\bar{n}=(1+\sqrt{\epsilon_{\rm Si}})/2$ is the average refractive index of the media above (air, neglecting the resist layer in the coupling to radiation) and below (Si) the metal layer, $\tilde{\alpha}$ is the ribbon polarizability per unit length, $S=4\pi^2/a\lambda_0\bar{n}$ describes radiative coupling, $a$ is the lattice period, $G=2\pi^2/3a^2\bar{\epsilon}+\ii S$ accounts for inter-ribbon interactions in the dipolar approximation, and $\bar{\epsilon}=(1+\epsilon_{\rm Si})/2$ is the average permittivity of the surrounding media. We express the polarizability
\[\tilde{\alpha}\approx-W^2\bar{\epsilon}\zeta_1^2\,\frac{1}{1/\eta_1+\ii\omega W\bar{\epsilon}/\sigma}\]
in terms of the 2D conductivity of the metal $\sigma$ and only consider the dominant contribution of the dipolar plasmon resonance corresponding to parameters \cite{paper303} $\eta_1\approx-0.0921+0.0233\,\ee^{-8.9\,d/W}$ and $\zeta_1\approx0.959-0.016\,\ee^{-39\,d/W}$, which depend on the ribbon thickness-to-width aspect ratio $d/W$. Finally, the 2D conductivity is related to the metal permittivity as $\sigma=(\ii\omega/4\pi)\left[(1-\epsilon_{\rm Ag})d+(1-\epsilon_{\rm c})d_{\rm c}\right]$, where we approximate the capping layer of thickness $d_{\rm c}=1.5\,$nm as an additional term in $\sigma$ with $\epsilon_{\rm c}=2$. We use tabulated optical data for silver \cite{JC1972} ($\epsilon_{\rm Ag}$) and crystalline silicon \cite{AS1983} ($\epsilon_{\rm Si}$). Reassuringly, the analytical theory just presented produces spectra in nearly full agreement with numerical electromagnetic simulations (see Figs.\ \ref{FigSI8} and \ref{FigSI9}). Incidentally, this analysis of ribbon arrays ignores the resist, which our numerical simulations (not shown) predict to only cause minor plasmon redshifts.

The transverse dipolar plasmon of a single ribbon is signaled by a divergence in $\tilde{\alpha}$ ({\it i.e.}, $\ii\omega\bar{\epsilon}/\sigma=-1/\eta_1W$), which combined with the dispersion relation of the extended film $\kpar=\ii\omega\bar{\epsilon}/2\pi\sigma\approx2\pi/\lamp$ leads to the condition
\[W=\frac{\lamp}{4\pi^2(-\eta_1)}\approx0.37\lamp\] for $d\ll W$.
Adopting this expression and neglecting inter-ribbon interactions, we can use Eq.\ (\ref{lpl0}) to readily obtain Eq.\ (\ref{l0W}) in the main text. It should be noted that inter-ribbon interaction can produce a small redshift correction in the plasmon position (see Fig.\ \ref{FigSI8}).

We find it convenient to arrange the above expressions above expressions by neglecting the capping layer and ap-
proximating the silver permittivity as $\epsilon_{\rm Ag}\approx1-\omega_{\rm bulk}^2/\omega(\omega+{\rm i}\gamma_{\rm in})$ in order to express the transmission coefficient of the array (Eq.\ (\ref{tarray})) as
\begin{align}
t=\frac{1}{\bar{n}}\left[1+\frac{\ii\omega\gamma_{\rm rad}}{\wp^2-\omega(\omega+\ii\gamma)}\right],
\nonumber
\end{align}
where
\begin{align}
\wp=\omega_{\rm bulk}\sqrt{\frac{1}{4\pi\bar{\epsilon}(-\eta_1)}\frac{d}{W}-\frac{\pi\zeta_1^2}{6\bar{\epsilon}}\frac{Wd}{a^2}}
\label{wparray}
\end{align}
is the resulting plasmon resonance of the array under normal incidence, whereas \[\gamma=\gamma_{\rm in}+\gamma_{\rm rad}\] is the total plasmon damping rate, contributed by the intrinsic component $\gamma_{\rm in}$ and a radiative component
\begin{align}
\gamma_{\rm rad}=\frac{\zeta_1^2}{2\bar{n}}\frac{\omega^2_{\rm bulk}}{c}\frac{Wd}{a}.
\label{gammarad}
\end{align}
The first term inside the square root of Eq.\ (\ref{wparray}) describes the plasmon frequency of the isolated ribbon, while the second term accounts for a redshift due to inter-ribbon interaction. We note that radiative damping (Eq.\ (\ref{gammarad})) decreases with increasing array period $a$, so sharper plasmons are expected in the limit of large separations, for which $\gamma\approx\gamma_{\rm in}$ (see Fig.\ \ref{Fig1}c); incidentally, we have neglected radiative contributions to the damping of individual ribbons under the assumption $W\ll\lambda_0$. When we specify Eq.\ (\ref{gammarad}) to Ag(111) ribbons on silicon, we find $\gamma_{\rm rad}=\Gamma\times Wd/a$, where $\hbar\Gamma=\zeta_1^2\hbar\omega^2_{\rm bulk}/(2\bar{n}c)\approx88\,$meV/nm.

\noindent {\bf Drude Damping Estimated from the Electrical Resistance.} We use the expression
\[\rho_{0,{\rm CGS}}[{\rm s}]\approx4\pi\times8.854\times10^{-12}\times\rho_{0,{\rm SI}}[\Omega\,{\rm m}]\]
to convert DC resistivities from SI to CGS units. Then, we use the Drude model to write the damping rate as
\[\gamma_{\rm in}=(4\pi)^{-1}\omega_{\rm bulk}^2\;\rho_{0,{\rm CGS}}.\]
Damping rates in Fig.\ \ref{Fig4}c are obtained by applying these formulas to the SI resistivities $\rho_{0,{\rm SI}}=1.62\times10^{-8}\,\Omega\,{\rm m}$ for bulk silver ($\gamma_{{\rm DC},N\gg1}$) and $\rho_{\rm S}Nd_{111}$ for silver films consisiting of $N$ Ag(111) atomic layers ($\gamma_{{\rm DC},N}$), where $d_{111}=0.236\,$nm is the atomic layer spacing and $\rho_{\rm S}$ is the average sheet resistance (for each value of $N$) obtained from the data points presented in Fig.\ \ref{Fig2}m.

\acknowledgments
\noindent We thank Marta Autore, Josep Canet-Ferrer, Rainer Hillenbrand, Johan Osmond, and Frederik Schiller for technical support and helpful discussions. V.M. and F.J.G.A. gratefully acknowledge generous help and hospitality from Luis Hueso and Ralph Gay at CIC nanoGUNE, where nanolithography and FTIR were performed. This work has been supported in part by ERC (Advanced Grant 789104-eNANO), the Spanish MINECO (MAT2017-88492-R, SEV2015-0522, and PCIN-2015-155, and MAT2016-78293-C6-6-R), the Catalan CERCA Program, the Basque Government (IT-1255-19), Fundaci\'o Privada Cellex, and the US National Science Foundation CAREER Award (1552461).

\clearpage
\section*{ADDITIONAL FIGURES}

\clearpage
%\bibliographystyle{apsrev}
%\bibliography{../../../bibtex/refs}

\end{document}